\documentclass[useAMS,usenatbib,usegraphicx]{mn2e}
\usepackage{bm}

\title[Accretion to Stars with Non-Dipole Magnetic Fields]{Accretion to Stars with Non-dipole Magnetic Fields}

\author[M. Long et al.]
{M. Long,$^1$\thanks{E-mail:Long@astro.cornell.edu}, M.M. Romanova,$^1$\thanks{romanova@astro.cornell.edu}, and R.V.E. Lovelace, $^{1,2}$\thanks{RVL1@cornell.edu}\\
$^1$ Department of Astronomy, Cornell University, Ithaca, NY 14853-6801, USA\\
$^2$ Department of Applied and Engineering Physics, Cornell University, Ithaca, NY 14853-6801, USA}

\begin{document}

%\date{Accepted . Received ; in original form }

\pagerange{\pageref{firstpage}--\pageref{lastpage}} \pubyear{2006}

\maketitle

\label{firstpage}

\begin{abstract}

Disc accretion to a rotating star with a non-dipole magnetic field is investigated for the first time in full
three-dimensional (3D) magnetohydrodynamic (MHD) simulations. We investigated the cases of (1) pure dipole, (2) pure
quadrupole, and (3) dipole plus quadrupole fields. The quadrupole magnetic moment $\mathbf{D}$ is taken to be parallel
to the dipole magnetic moment $\bm{\mu}$, and both are inclined relative to the spin axis of the star $\bm{\Omega}$ at
an angle $\Theta$. Simulations have shown that in each case the structure of the funnel streams and associated hot
spots on the surface of the star have specific features connected with the magnetic field configuration. In the pure
dipole case matter accretes in two funnel streams which form two arch-like spots near the magnetic poles. In the case
of a pure quadrupole field, most of the matter flows through the quadrupole ``belt" forming a ring-shaped hot region on
the magnetic equator. In the case of a dipole plus quadrupole field, magnetic flux in the northern magnetic hemisphere
is larger than that in the southern, and the quadrupole belt and the ring are displaced to the south. The stronger the
quadrupole, the closer the ring is to the magnetic equator. At sufficiently large $\Theta$, matter also flows to the
south pole, forming a hot spot near the pole. The light curves have a variety of different features which makes it
difficult to derive the magnetic field configuration from the light curves. There are specific features which are
different in cases of dipole and quadrupole dominated magnetic field: (1) Angular momentum flow between the star and
disc is more efficient in the case of the dipole field; (2) Hot spots are hotter and brighter in case of the dipole
field because the matter accelerates over a longer distance compared with the flow in a quadrupole case.

\end{abstract}


\section{Introduction}

Many accreting stars have a strong magnetic field which disrupts the accretion disc and channels accreting matter to
the star's surface. For such stars many spectral and photometric properties are determined by the magnetic field.
Examples include young solar-type stars (Classical T Tauri stars, hereafter  CTTS, e.g., Hartmann 1994), neutron stars
in binary systems (X-ray pulsars and millisecond pulsars, e.g. Ghosh \& Lamb 1978; Chakrabarty et al. 2003), white
dwarfs in the binary systems (cataclysmic variables, e.g. Wickramasinghe et al. 1991; Warner 1995, 2000), and also
brown dwarfs (e.g., Scholz \& Ray 2006). The properties of such accretion will depend on the structure of the magnetic
field of the star.

In early models it was assumed that the star's intrinsic magnetic field was a pure {\it dipole} (Ghosh \& Lamb 1979a,b;
Camenzind 1990; K\"onigl 1991). Recently, 2D and full 3D MHD numerical simulations of magnetospheric accretion  were
performed which confirmed many predicted features and revealed many new details of accretion to a star with a dipole
field (Romanova et al. 2002, 2003, 2004; Long et al. 2005; Kulkarni \& Romanova 2005). Different aspects of the
disc-magnetosphere interaction were also investigated numerically (Hayashi et al. 1996; Hirose et al. 1997; Miller, \&
Stone 1997; Goodson, Winglee \& B\"{o}hm 1997, 1999; von Rekowski \& Brandenburg 2004, 2006) and theoretically (e.g.,
Lovelace, Romanova \& Bisnovatyi-Kogan 1995; Uzdensky, K\"onigl \& Litwin 2002;  Matt \& Pudritz 2004, 2005).

However, the actual configuration of the magnetic field of strongly magnetized stars may depart from the dipole one.
For example, Safier (1998) presented a number of arguments pointing to a non-dipolar magnetic fields in some CTTSs.
The Zeeman measurements of the magnetic field of a number of CTTSs based on the photospheric lines show that the
magnetic field at the surface of CTTSs is strong (1-3 kGs) but not ordered, which means that close to the star the
magnetic field is non-dipole (e.g., Johns-Krull et al. 1999; Johns-Krull,  Valenti, \& Koresko 1999; Johns-Krull \&
Gafford 2002; Smirnov et al. 2003).

The measurements of the magnetic fields of rapidly rotating low-mass stars with the Zeeman-Doppler imaging technique
have shown that in a number of stars the magnetic field has a complicated multipolar topology close to the star (Donati
\& Cameron 1997; Donati et al. 1999; Jardine et al. 2002). If the multipolar component dominates in the disc-accreting
binary systems, then  the matter flow to the star and the hot spots will be different from those in the case of a pure
dipole field (e.g., Jardine et al. 2006).

The dipole magnetic field may of course dominate in some stars, and it will dominate at larger distances from stars
with complex surface magnetic fields. There are observational signs that the dipole component possibly disrupts the
disc. In many cases there is evidence that the dipole field dominates at all distances, giving  periodic light curves
in X-ray binaries, intermediate polars, and some CTTSs. In addition, direct measurements of the magnetic field using
polarization of some spectral lines shows evidence of the pure dipole field in a number of the CTTSs (Valenti \&
Johns-Krull 2004; Symington et al. 2005). An remarkable result was recently obtained by Donati et al. (2006) who
recovered the magnetic field of the convective low- mass star v374Peg and obtained a pure dipole configuration. Thus
the magnetic field configuration is likely to be different in stars with varying levels of importance of the dipole
component. Bouvier et al. (2006) argued that there are many signs of magnetospheric accretion in CTTSs and that the
field is probably dipolar at larger distances from the star, but may have a strong multipolar component close to the
star.

Recently, the first theoretical and numerical research was done on accretion to a star with a non-dipole magnetic
field. Jardine et al. (2006) investigated the possible paths of the accreting matter in the case of a multipolar field
derived from observations. Donati et al. (2006) developed a simplified stationary model for such accretion. von
Rekowski and Brandenburg (2006) performed axisymmetric simulations of the disc-magnetosphere interaction in case when
magnetic field is generated by the dynamo processes both, in the star and in the disc. They obtained a time-variable
magnetic field of the star with a complicated multipolar configuration which  shows that the dynamo may be responsible
for a complex magnetic field structure.

In this paper we show the first results of full 3D MHD simulations of accretion to a star with a  non-dipole magnetic
field frozen to the stellar surface. As a first step, we take a combination of the dipole and quadrupole fields with
aligned axes and investigate properties of the funnel streams and hot spots at the surface of the star. More
complicated structures of the field will be considered in future simulations. Analytical analysis of accretion to a
non-rotating star with a pure quadrupole magnetic field was done by Lipunov (1978). Combination of dipole and multipole
fields was discussed by Lovelace et al. (2005). However, no numerical simulations of such accretion have been performed
so far.

In \S 2 we describe our numerical model and the  magnetic field configurations used in the simulations. We describe our
simulation results in \S 3, \S 4 and study the dependence on different parameters in \S 5 and give a summary of our
work in \S 6.

\section{Numerical Model and Initial Magnetic Fields}

The numerical model used in this paper has been described in a number of previous papers (e.g. Koldoba et al. 2002;
Romanova et al. 2002, 2004; Ustyugova et al. 2006). We describe briefly the main aspects of the numerical model in the
Appendix. In this section, we describe the initial conditions, boundary conditions, and new configurations of the
dipole plus quadrupole field used in this paper.

\subsection{Initial Conditions}

The region considered consists of the star located in the center of coordinate system, a dense disc located in the
equatorial plane and a low-density corona which occupies the rest of the simulation region. Initially, the disc and
corona are in {\it rotational} hydrodynamic equilibrium. That is, the sum of the gravitational, centrifugal, and
pressure gradient forces is zero at each point of the simulation region. The initial magnetic field is a combination of
dipole and a quadrupole field components which are force-free at $t=0$. The initial rotational velocity in the disc is
close to Keplerian (i.e., the small pressure component is taken into account). The corona at different cylindrical
radii $r$ rotates with angular velocities corresponding to Keplerian velocity of the disc at this distance $r$. This
initial rotation is assumed so as to avoid a strong initial discontinuity of the magnetic field at the boundary between
the  disc and corona. The initial rotation of the disc with a non-rotating corona leads to a strong magnetic forces at
the disc-magnetosphere boundary and to a strong initial torque of the disc. The distribution of density and pressure in
the disc and corona and the complete description of these initial conditions is given in Romanova et al. (2002) and
Ustyugova et al. (2006).

The initial accretion disc extends inward to an inner radius $r_d$ and has a temperature $T_d$ which is much less than
the corona temperature $T_d=0.01T_c$. The density of the disc is 100 times the density of the corona,
$\rho_c=0.01\rho_d$. The values $T_c,T_d,\rho_c,\rho_d$ are determined at the boundary between the disc and corona near
the inner radius of the disc.

After the beginning of the rotation, the magnetic field lines start to twist and the exact force-balance is disturbed
because the magnetic forces begin to act. However, these forces are not strong enough to disturb the disc
significantly. The accretion rate  increases only slightly as a result of the magnetic braking associated with this
initial twist. These initial conditions allow an  investigation of accretion to a star with a  dipole magnetic field
for conditions where the disc matter accretes inward very slowly on the viscous time-scale. The $\alpha-$ viscosity
incorporated in our code permits the regulation of the accretion rate. Earlier, we used these initial conditions to
investigate magnetospheric flow in case of a pure  dipole intrinsic field (Romanova et al. 2002, 2004).

\subsection{Boundary Conditions}

At the inner boundary ($R=R_{*}$), ``free" boundary conditions are applied for the density, pressure, entropy, velocity
and $\phi-$component of the magnetic field, ${\partial\rho}/{\partial R}=0$, ${\partial p}/{\partial R}=0$, ${\partial
S}/{\partial R}=0$, ${\partial ({\bf v} - {\bf \Omega}\times {\bf R})_R}/ {\partial R}=0$, ${\partial(R
B_\phi)}/{\partial R}=0$. The magnetic field is frozen to the surface of the star, so that the magnetic flux
$\Psi(R,\theta)$, at the inner boundary is derived from the frozen-in condition ${{\partial \Psi}/{\partial t}} + {\bf
v}_p\cdot{\bf \nabla}\Psi = 0$. The poloidal components $B_R$ and $B_\theta$ are derived from the magnetic flux
function $\Psi(R,\theta)$.

At the outer boundary, free boundary conditions are taken for all variables. Matter does not  inflow through the outer
boundary. The investigated numerical region is large, $\sim 45 R_*$, so that the  initial reservoir of matter in the
disc is large and sufficient for the performed simulations. Matter flows slowly inward from external regions of the
disc. During the  simulation times studied here only a very small fraction of the total disc matter accretes to the
star. Test simulations were done where matter was allowed to accrete inward from an external boundary did not change
the result.

\subsection{Reference Units}

Dimensionless reference values are used in our simulations, which allows our models to apply to a range of different
systems. We let $R_0$ denote a reference distance scale, which is equal to $R_*/0.35$, where $R_*$ is the radius of the
star. The subscript ``0" denotes reference values. We take the reference values for the velocity, angular speed and
timescale to be $v_0=(GM/R_0)^{1/2}$, $\Omega_0=v_0/R_0$ and $P_0=2\pi R_0/v_0 $ respectively. The reference magnetic
field $B_0=B_*(R_*/R_0)^3$, where $B_*$ is magnetic field at the surface of the star. The reference dipole magnetic
moment $\mu_0=B_0R_0^3$, and the reference quadrupole moment $D_0=B_0R_0^4$. The reference density, pressure and mass
accretion rate are $\rho_0=B_0^2/v_0^2$, $p_0=\rho_0v_0^2$ and $\dot{M} _0=\rho_0v_0R_0^2$ respectively. The
dimensionless variables are $\widetilde{R}=R/R_0$, $\widetilde{v}=v/v_0$, time $\widetilde{t}=t/P_0$ etc. In the
following discussion these dimensionless units are used but the tildes are implicit. The grid is $N_R=75$, $N=31$,
which gives our external boundary at $R_{max}\approx 16$.

For a typical CCTS, we take the stellar mass $M_*=0.8M_{\odot}$, its radius $R_*=1.8R_{\odot}$ and the magnetic field
at the surface of the star $B_*=1$kG, and thus obtain the reference units discussed above. Therefore, the case of
$\mu=0.5$ and $D=0.5$ investigated in the following sections corresponds to the magnetic field with the dipole
component $B_{*d}=500$G and quadrupole component $B_{*q}=500$G.

\subsection{Magnetic Fields of the Star}

\begin{figure}
\includegraphics[scale=0.5, angle=0]{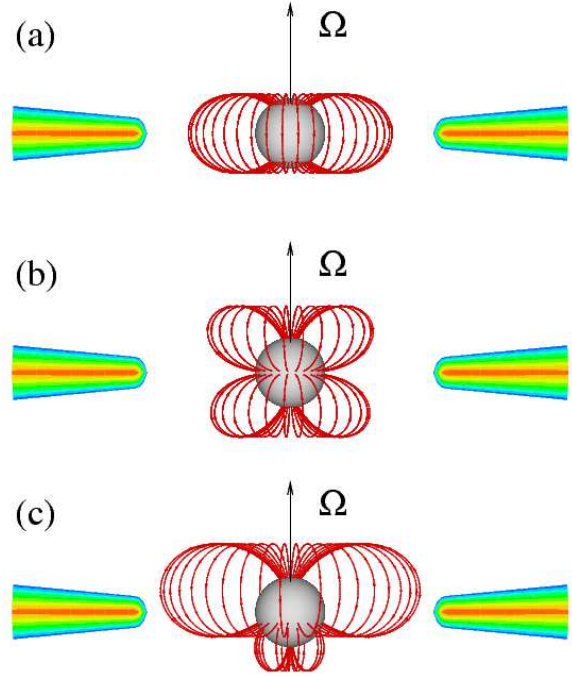}
\caption{\label{fig1} Initial configurations of the magnetic fields used in simulations.  (a) Pure dipole field with
$\mu\neq0$, $D=0$; (b) Pure quadrupole field $D\neq 0$; (c) Dipole plus quadrupole field, $\mu\neq0$, $D\neq0$. The
inner region of the accretion disc is also seen in the figures.}
\end{figure}

In the absence of plasma outside the star we would have $\nabla\times\mathbf{B}=0$, so that the intrinsic magnetic
field of the star can be written as $\mathbf{B}=-\nabla\varphi$, where the scalar potential of the magnetic field is
$\varphi(\mathbf{r})= \Sigma{m_a}/{|\mathbf{r-r}_a|}$, and $m_a$ is an analogy of the magnetic ``charge", $\mathbf{r}$
and $\mathbf{r}_a$ are the positions of the observer and the magnetic ``charges" respectively. The scalar potential can
be represented as a multipole expansion in powers of $1/r$ (e.g. Landau \& Lifshitz 1983),
\begin{equation}
\varphi=\Sigma\frac{m_a}{|\mathbf{r-r}_a|} =\varphi^{(0)}+\varphi^{(1)}+\varphi^ {(2)}+...  \ \ .
\end{equation}
The first term $\varphi^{(0)} =0$ because there are no monopoles. The second term is the dipole component,
$\varphi^{(1)}=\mu\cos\theta/{r^2}$, where $\mu$ is the dipole moment. The third term is the quadrupole component,
\[
\varphi^{(2)}=\frac{D_{\alpha\beta}}{6}\frac {\partial^2}{\partial x_{\alpha}\partial
x_{\beta}}\left(\frac{1}{r}\right)~,
\]
where $x_\alpha$, $x_\alpha$ are the components of $\mathbf{r}$, $D_ {\alpha\beta}$ is the magnetic quadrupole moment
tensor, and the sum over repeated indices is implied. Because $D_{\alpha\alpha}=0$, we have, $\varphi^
{(2)}={D_{\alpha\beta}n_{\alpha}n_{\beta}}/{2r^3}$, where $n_{\alpha}=x_\alpha/r$, $n_{\beta}=x_\beta/r$.

Considering the \textit{axisymmetric} case where $D_{11}=D_{22}=-D_{33}/2$, we call $D=D_{33}$ the quadrupole moment
and refer to the axis of symmetry as the ``direction'' of the quadrupole moment. In spherical coordinates
$(r,\theta,\phi)$, we have,
\begin{equation}
\varphi^{(2)}=\frac{D}{4r^3}(3\cos^2\theta-1)=\frac{D}{2r^3}P_2(\cos\theta) \ \ ,
\end{equation}
where $P_2$ is a Legendre polynomial. Then we can get the quadrupole magnetic field $\mathbf{B}$ and the flux function
$\Psi$:

\begin{eqnarray}
B_r&\equiv&-\frac{\partial\varphi}{\partial r}=\frac{3D}{4r^4}(3\cos^2\theta-1) \ \ , \\
B_\theta&\equiv&-\frac{1}{r}\frac{\partial\varphi}{\partial\theta}=\frac{3D}
{2r^4}\sin\theta\cos\theta \ \ , \\
\Psi&=&\frac{3D}{4r^2}\sin^2\theta\cos\theta \ \ .
\end{eqnarray}

In our simulations we use a combination of dipole and quadrupole magnetic fields:
\begin{eqnarray}
B_r&=&\frac{2\mu\cos\theta}{r^3}+\frac{3D}{4r^4}(3\cos^2\theta-1) \ \ , \\
B_\theta&=&\frac{\mu\sin\theta}{r^3}+\frac{3D}{2r^4}\sin\theta\cos\theta \ \ ,\\
\Psi&=&\frac{\mu\sin^2\theta}{r}+\frac{3D}{4r^2}\sin^2\theta\cos\theta \ \ .
\end{eqnarray}

The star has fixed dipole and/or quadrupole magnetic fields with moments $\mu$ and $D$ respectively. For simplicity,
they are in the same direction and are inclined with respect to the rotational axis $\Omega$ at an angle $\Theta$. We
investigated accretion at angles $\Theta=0^\circ$, $\Theta=30^\circ$, and $\Theta=60^\circ$. Spin axes of the star and
the disc coincide.

We vary the value of the quadrupole moment from a pure dipole case, $D=0$, up to a pure quadrupole case, $\mu=0$.
Figure 1 shows the main initial configurations of the magnetic field considered in the paper: (a) pure dipole
configuration, $D=0$; (b) pure quadrupole case $\mu=0$;  and (c) the case when the dipole and quadrupole fields are of
the same order. One can see that when both dipole and quadrupole fields are present, the field is not symmetric
relative to the equatorial plane: the magnetic flux $\Psi$ is larger in the northern hemisphere where the dipole field
is added to the quadrupole field, and smaller in the southern hemisphere where the dipole field is subtracted from the
quadrupole field. Simulations have shown that this feature of the dipole plus quadrupole field leads to results
different from the pure dipole case.

\section{Results of Simulations}

\subsection{Accretion to a Star with Pure Dipole and Pure Quadrupole Magnetic
Fields}

\begin{figure*}
\begin{center}
\includegraphics[angle=0,width=11cm]{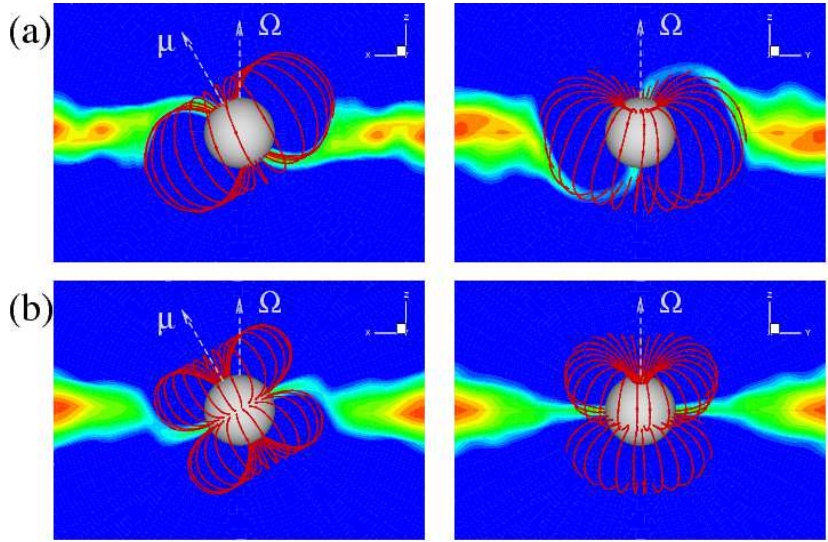}
\caption{\label{fig2} Disc accretion to a star with (a) pure dipole and (b) pure quadrupole magnetic configurations at
$\Theta=30^\circ$ and at time $t=5$. The left panels show projections in the $x-z$ ($\mu\Omega$) plane, and the right
panels show projections in the $y-z$ plane. The background shows density contours, and the red lines show the outermost
closed field lines. }
\end{center}
\end{figure*}

First, for reference we show the results of simulations of accretion to a star with a pure dipole field with $\mu=0.5$
inclined relative to the spin axis at an angle $\Theta=30^\circ$. Figure 2a (left panel) shows the density distribution
in the $x-z$ plane (this is also the $\bm\mu-\mathbf{\Omega}$ plane). One can see that most of the matter flows through
symmetric funnel streams to the nearest magnetic pole. A smaller amount of matter flows to the opposite pole. The right
panel shows the cross- section in the $y-z$ plane, where similar but weaker streams are observed. In fact at
sufficiently low density levels the matter blankets the magnetosphere completely (Romanova et al. 2003) so that at
these levels any cross-section will show the funnel streams.

Figure 2b shows the results of accretion to a pure quadrupole magnetic configuration with $\mu=0$, and $D=0.5$, which
is inclined at an angle $\Theta=30^\circ$. The left panel shows that in the $x-z$ cross-section, matter is slightly
lifted above the equatorial plane and flows through the quadrupole belt to the equatorial region of the star. The right
panel shows that in the $y-z$ plane, matter flows straight to the equatorial belt without being lifted.

Figure 3a shows the three-dimensional distribution of matter flow around a star with a pure quadrupole field, where one
of density levels is shown in green color. The left projection demonstrates that the matter flows in  a thin sheet
approximately in the plane of the magnetic equator. The right panel shows that this flow is wide and is not
axisymmetric. Thus, in the pure quadrupole case matter flows to the star in a thin, wide sheet in the region of the
magnetic equator.

\begin{figure*}
\begin{center}
\includegraphics[angle=0,width=11cm]{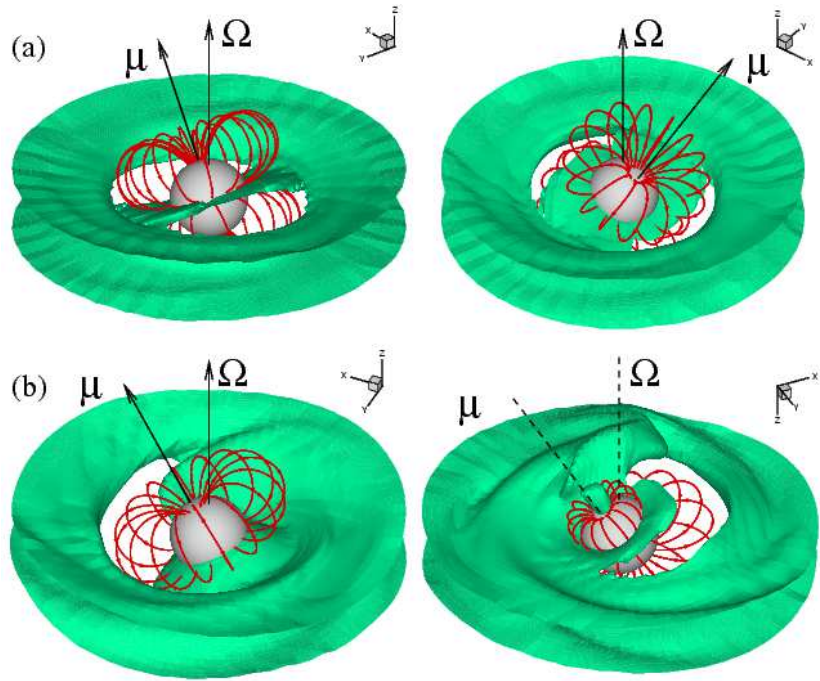}
\caption{\label{fig3} Three-dimensional picture of matter flow around the star for the pure quadrupole case (top panels
-a ) and for a dipole plus quadrupole case at $\mu=0.5$, $D=0.5$ (bottom panels -b ). In both cases $\Theta=30^\circ$.
The background shows one of the density levels, $\rho=0.2$. Red lines are the outermost closed field lines. In (b), the
dashed lines mean the south part of the rotation axis and magnetic axis.}
\end{center}
\end{figure*}

\subsection{Accretion to a Star with a Dipole plus Quadrupole Magnetic Field}

\begin{figure*}
\begin{center}
\includegraphics[angle=0,width=11cm]{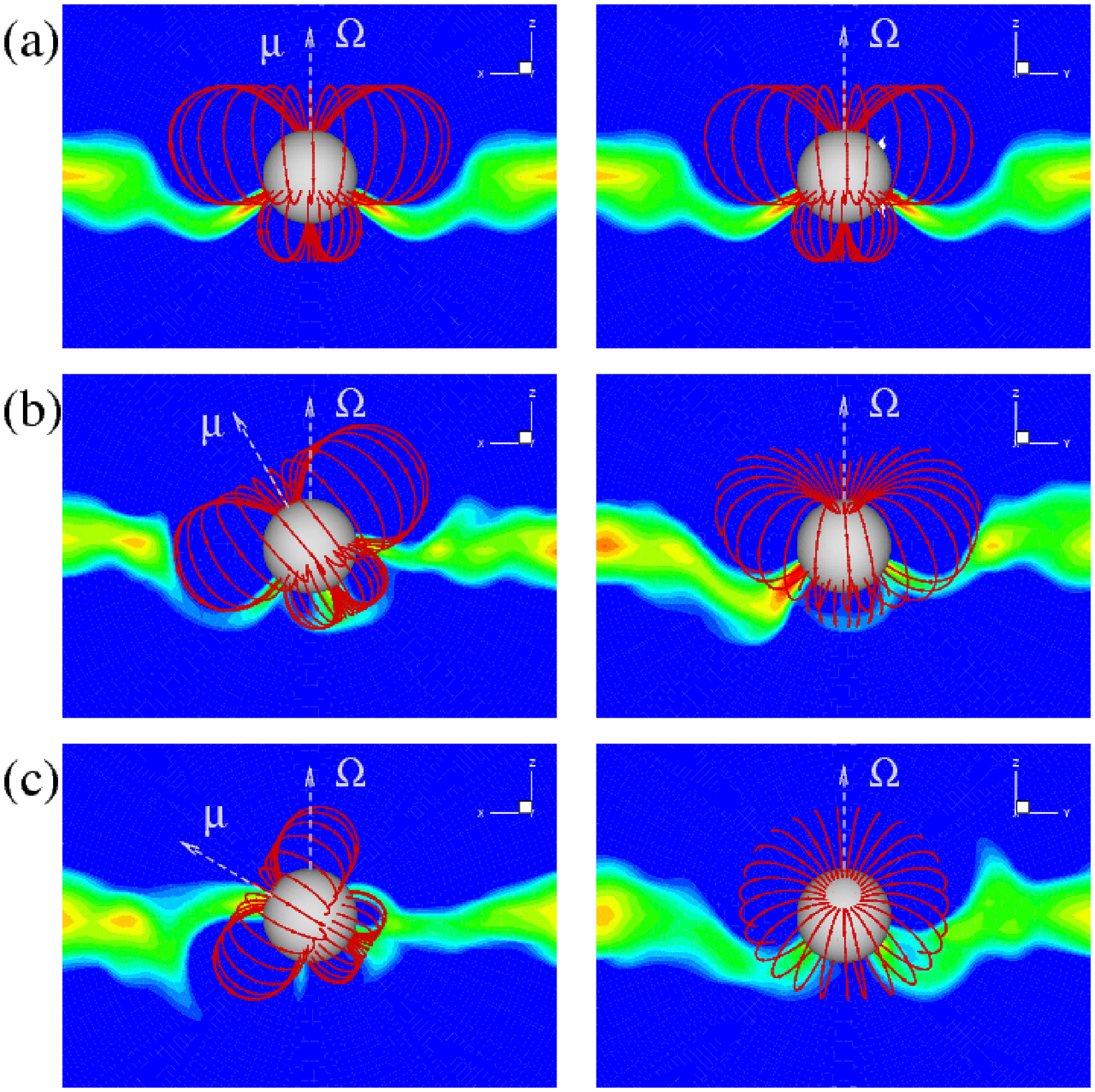}
\caption{\label{fig4} Disc accretion to a star with a dipole plus quadrupole magnetic field $\mu=0.5$, $D=0.5$ for
different misalignment angles: (a) $\Theta=0^\circ$; (b) $\Theta=30^\circ$; (c) $\Theta=60^\circ$ at time $t=5$. The
left panels show projections in the $x-z$ ($\mu\Omega$) plane, and the right panels show projections in the $y- z$
plane. The background shows density contours, and the red lines show the outermost closed field lines. }
\end{center}
\end{figure*}

\begin{figure*}
\begin{center}
\includegraphics[angle=0,width=10cm]{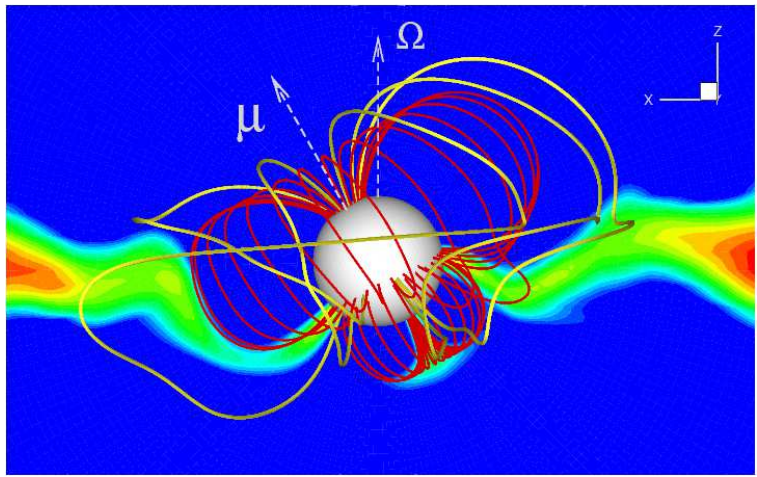}
\caption{\label{fig5} The detailed structure of disc accretion to a star with a dipole plus quadrupole field, with
$\mu=0.5$, $D=0.5$, $\Theta=30^\circ$. In addition to the closed field lines (red lines), a number of sample open field
lines (yellow lines) are shown. }
\end{center}
\end{figure*}

\begin{figure*}
\includegraphics[angle=0,width=11cm]{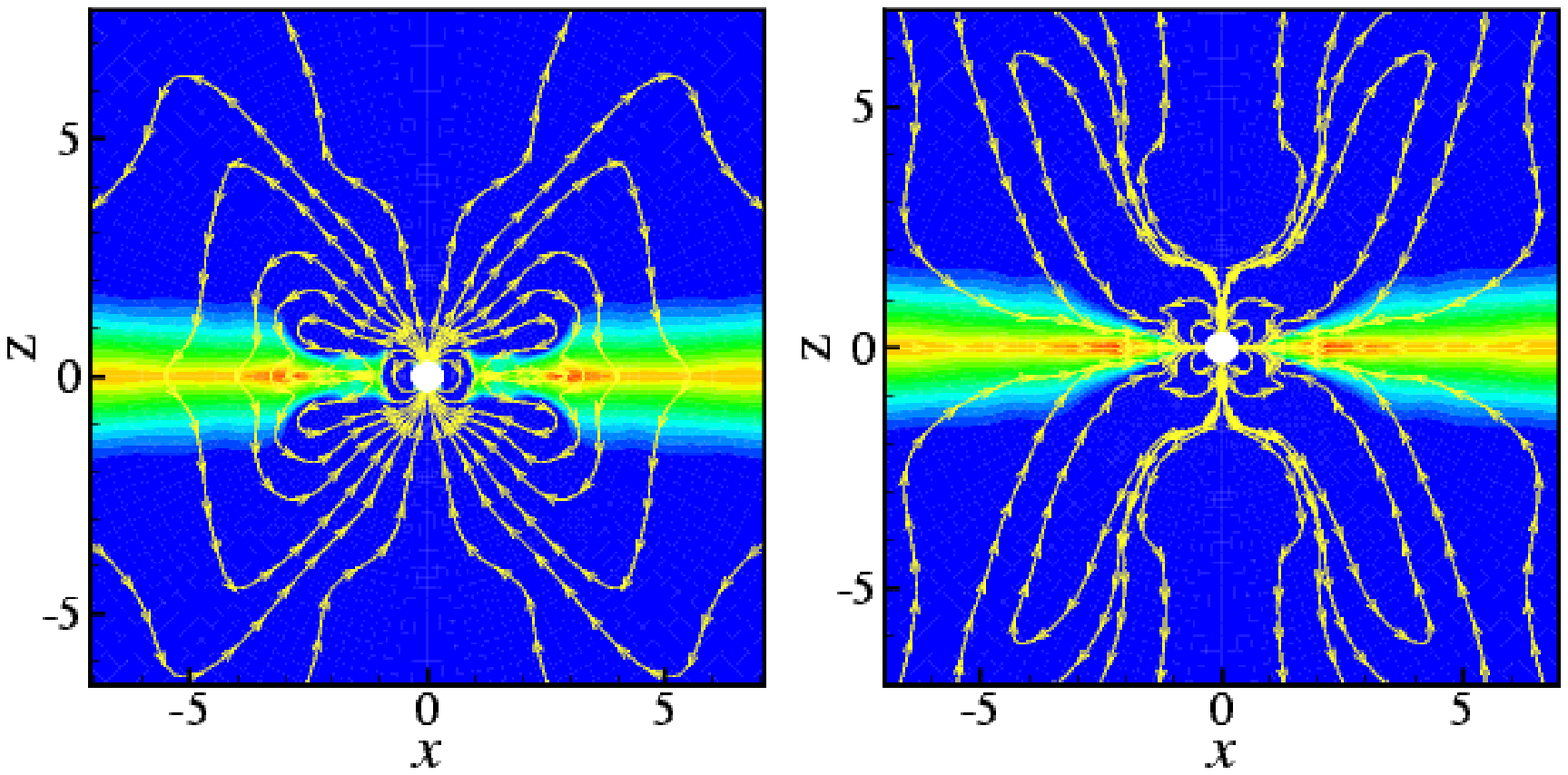}
\caption{\label{fig6} The large-scale structure of the magnetic field (yellow lines) in the cases of pure dipole at
$\mu=0.5$ (left panel) and pure quadrupole $D=0.5$ (right panel) at $\Theta=0$. The background shows the density
distribution where density varies from $\rho=2$ (red) to $\rho=0.01$ (blue). }
\end{figure*}

We calculated the case of the dipole plus quadrupole magnetic field at three misalignment angles, $\Theta=0^\circ$,
$\Theta=30^\circ$, and $\Theta=60^\circ$. In all cases we took $\mu=0.5$ and $D=0.5$. This magnetic configuration is
not symmetric relative to the equatorial plane. In the northern hemisphere the magnetic flux of the dipole is added to
that of the quadrupole, so that magnetic flux is larger than the pure dipole one. In the southern hemisphere, the
magnetic flux of the dipole is subtracted from that of the quadrupole, so that the flux is smaller than the dipole
flux.

Figure 4a shows the $x-z$ (left panels) and $y-z$ (right panels) cross-sections of accretion to a star with
$\Theta=0^\circ$. One can see that the matter flows axisymmetrically to the quadrupole equatorial belt, which is now
located in the southern hemisphere.

Figure 4b shows the case $\Theta=30^\circ$. Here, part of the matter flows to the south polar region. Figure 3b shows
the 3D view of such a flow in the northern hemisphere. One can see that matter flows through the quadrupole belt, but
the flow is not symmetric relative to the $z-$axis. The right panel shows the flow as seen from the southern
hemisphere, where part of the matter flows to the south pole region.

Figure 4c shows that at $\Theta=60^\circ$, the flow is even more complicated: some matter flows to the north pole, some
matter flows to the south pole, and some matter flows through the belt.

So, one can expect that in the case of $\Theta\approx 0^\circ$, accretion to a quadrupole or a combined
dipole-quadrupole field will go through a quadrupole belt and will form a ring-like spot at the surface of the star.
However, at large misalignment angles $\Theta$ the structure of the accreting streams is more complicated.

\subsection{Disc-Magnetosphere Interaction and Magnetospheric Radius}

Here we chose one of typical cases with mixed dipole and quadrupole fields,  $\mu=0.5$, $D=0.5$, and $\Theta=30^\circ$,
and consider it in greater detail. First, we estimate the modified plasma $\beta$ parameter,
\[\beta=\frac{p+\rho v^2}{B^2/8\pi} \approx \frac{\rho v_\phi^2}{B^2/8\pi}, \]
where we took into account that the ram pressure of matter in the disc is much larger than the thermal pressure. The
surface $\beta=1$ separates the regions of magnetically dominated and matter dominated plasma. This surface crosses the
equatorial plane at the magnetospheric radius $r_m$, which is $r_m\approx1.14 $. One can see from Figure 5 that this is
the radius where the disc is disrupted and matter goes to the funnel streams. Both dipole and quadrupole components
contribute to the magnetosphere and to the disruption of the disc.

For $r<r_m$, the magnetic energy-density dominates, and the field lines are closed. However, at larger radii the field
lines are dragged by the disc and also inflate and open in the corona. Figure 5 shows the outermost closed field lines
(red lines) and also some open field lines (yellow lines) in the corona.

%%%%added%%%%
Opening and inflation of the field lines is somewhat different between the dipole-dominated and quadrupole-dominated
fields. Figure 6 shows the large-scale structure of the magnetic field in the dipole case at $\mu=0.5$, $D=0$ (left
panel) and in the quadrupole case at $D=0.5$, $\mu=0$ (right panel). One can see that in both cases there is a closed
magnetosphere close to the star and a set of inflated field lines at larger distance. We should note that in our 3D
simulations, inflation is somewhat suppressed by relatively dense corona and in reality it may be more efficient. From
the other side, these plots are shown for $t=8$ and at longer time inflation of the external field lines may progress
further. In spite of that, the typical pattern of the inflated field lines is different for the dipole and for the
quadrupole. In both cases the open field lines participate in the spin-down of the star (see Section 5).

%%%%%%%%

\section{Hot Spots and Light Curves}

\begin{figure*}
\begin{center}
\includegraphics[angle=0,width=12cm]{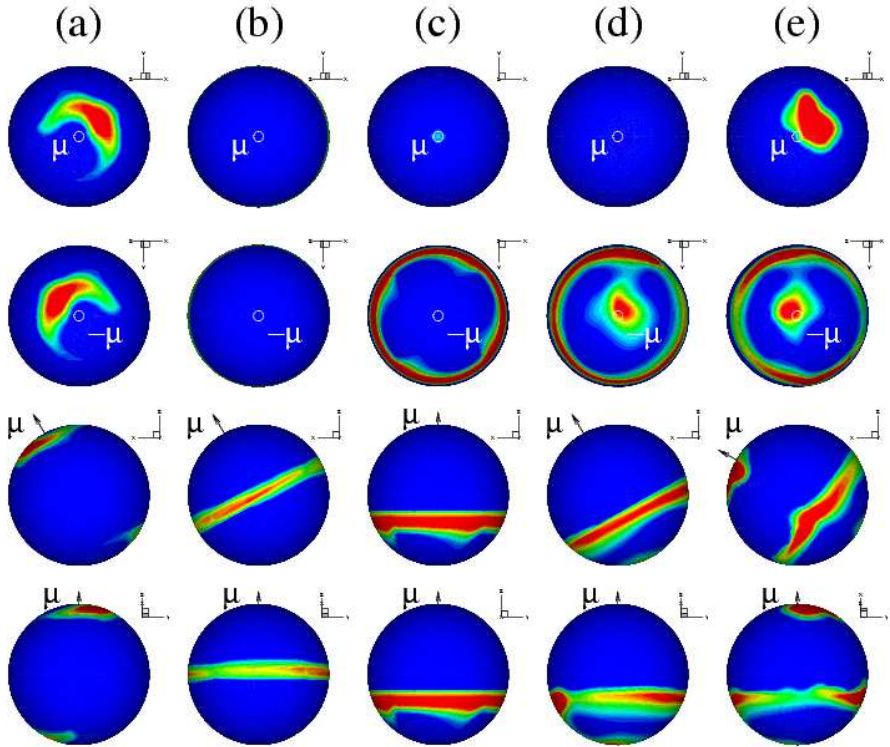}
\caption{\label{fig7} The hot spots at t=5 shown at different projections: from the top to the bottom, each column
shows views from the north and south magnetic poles, the $x-z$ projection and the $\mu-y$ projection respectively. The
background shows density contours. Red color corresponds to the maximum density ($\rho=2.5$) and dark blue corresponds
to the minimum density $\rho=0.01$. (a) pure dipole case, $\mu=0.5$, $D=0$ at $\Theta=30^\circ$; (b) pure quadrupole
case, $\mu=0$, $D=0.5$ at $\Theta=0^\circ$. Columns (c), (d) and (e) show spots for the $\mu=D=0.5$ case for
$\Theta=0^\circ$, $\Theta=30^\circ$ and $\Theta=60^\circ$ respectively.}
\end{center}
\end{figure*}

The inflowing matter of the funnel streams falls onto the star and forms hot spots on the stellar surface. The shape of
the hot spots reflects the shape of the funnel streams. Earlier 3D simulations of accretion to a star with a misaligned
dipole magnetic field have shown that hot spots are always similar in the northern and southern hemispheres (Romanova
et al. 2004). Our  present simulations with a dipole plus quadrupole magnetic fields have shown that hot spots are
always \textit{different} in the northern and southern hemispheres.

Figure 7 shows the hot spots for different magnetic configurations at $t=5$. From the top to the bottom, each column
shows view from the north and south magnetic poles, the $x-z$ projection and the $\mu-y$ projection respectively.

Column (a) represents the results for a pure dipole magnetic field with $\mu=0.5$ and misalignment angle
$\Theta=30^\circ$. One can see that there are two hot spots near the magnetic poles which are similar to each other.
Column (b) shows the results for a pure quadrupole configuration at $\Theta=30^\circ$. The inflowing matter forms a
ring near the magnetic equatorial plane in the region of the quadrupole belt. The ring has two density enhancements
from two opposite sides because the flow is not axisymmetric (See also Figure 3a). Columns (c), (d) and (e) show the
results for a dipole plus quadrupole field, $\mu=0.5$, $D=0.5$, for misalignment angles $\Theta=0$, $\Theta=30^\circ$
and $\Theta=60^\circ$ respectively. In the case of $\Theta=0^\circ$, (column c), the hot spot has the shape of the ring
located below the equatorial plane. The ring is below the equator because the magnetic field in the northern hemisphere
is much stronger than that in the southern hemisphere and the quadrupole belt is displaced to the south. The ring is
symmetric relative to $\mathbf{\Omega}$ axis because $\Theta=0^\circ$. There are no hot spots near the poles (see also
Figure 4a). Column (d) shows the case when $\Theta=30^\circ$. One can see that the magnetic flux in northern hemisphere
is still strong enough to stop the flow to the north magnetic pole, so there is no hot spot. However, some matter flows
to the south magnetic pole along closed field lines of the southern part of the magnetosphere. Column (e) shows the
results for $\Theta=60^\circ$. Now the magnetic axis is tilted more towards the disc, which leads to matter flow to
both the north and south magnetic poles. We can still see the belt in the southern hemisphere.

\begin{figure}
\includegraphics[scale=0.5, angle=0]{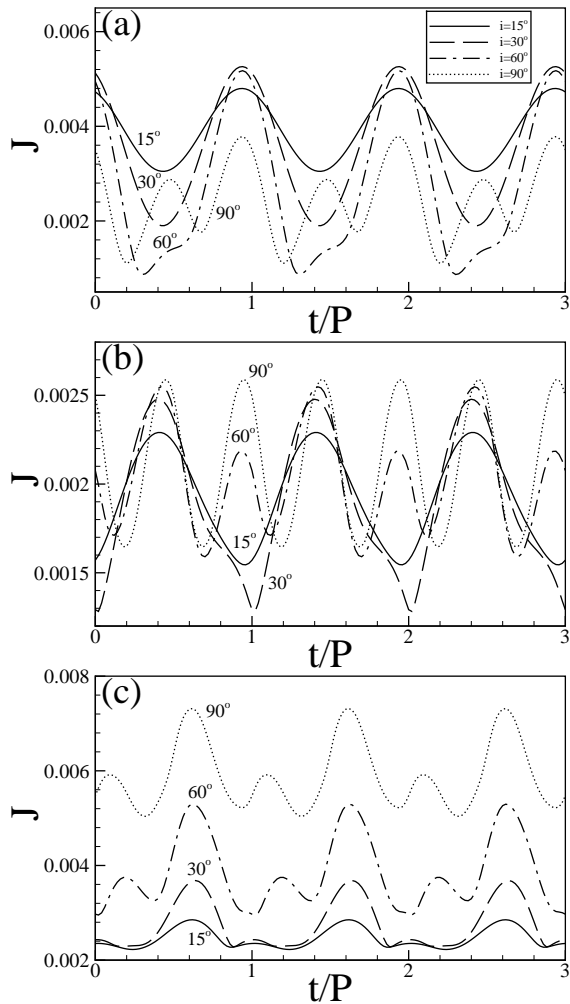}
\caption{\label{fig8} Light curves for different magnetic field configurations at misalignment angle $\Theta=30^\circ$
for different inclination angles $i$. The inclination angle is measured relative to the star's rotation axis. Solid,
dash, dash-dot and dotted lines represent the results for $i=15^\circ$, $i=30^\circ$, $i=60^\circ$, $i=90^\circ$
respectively. (a) Pure dipole case $\mu=0.5$, $D=0$; (b) Pure quadrupole case $\mu=0$, $D=0.5$; (c) dipole plus
quadrupole case at $\mu=0.5$, $D=0.5$. }
\end{figure}

Consider now the light curves  from  hot spots on the star for different magnetic configurations. We assume that the
kinetic energy of the inflowing matter is converted into black-body radiation and is radiated isotropically (e.g.
Romanova et al. 2004). The variation in the light curve is associated with the position and shape of the hot spots.
This means that the evolution of the hot spots can affect the light curves. After several rotations of the star, the
shape and position of hot spots do not vary significantly. Therefore, as a first approximation we calculate the light
curves by fixing the hot spots at $t=5$ and rotating the star.

The observed intensity of radiation in the direction $\mathbf{\hat{k}}$ is $J=\int f(\mathbf{R},\mathbf{k})
\mathrm{d}S$, where $f(\mathbf{R},\mathbf{\hat{k}})$ is the intensity of the radiation from a unit area into the solid
angle element $d\Omega$ in the direction $\mathbf{\hat{k}}$, $dS$ is an element of the surface area of the star,
$\mathbf{R}$ is the radius vector. It can be calculated from our simulations (see details in Romanova et al. 2004).

Figure 8 shows the light curves for a pure dipole, pure quadrupole and a dipole plus quadrupole case at
$\Theta=30^\circ$ at different inclination angles from $i=15^\circ$ to $90^\circ$. The inclination angles are measured
relative to the star's rotation axis. One can see that for the pure dipole case (top panel) the observer will see only
one peak per period, excluding the case of very high inclination angles, $i\geq70^\circ$. In the pure quadrupole case
(middle panel), the observer will see two peaks per period starting from smaller inclination angle, $i\geq40^\circ$. In
the case of the mixed dipole plus quadrupole field, there are also two peaks per period at $i\geq40^\circ$. Also, when
the inclination angle is changed, the change in amplitude is stronger in this case. If the inclination angle of the
star is known, one can derive the magnetic configurations by varying $\Theta$. However, there may be several possible
configurations of the field and $\Theta$ which give similar light curves.

\section{Dependence on the Strength of the Quadrupole Component}

Here, we consider the dependence of different features of the accretion flow and hot spots on the strength of the
quadrupole component. We fixed the dipole component at $\mu=0.5$ and varied the quadrupole component between $D=0$ and
$D=2.5$. First, we calculated the mass accretion rate $\dot{M}=-\int d\bm{S} \cdot\rho\bm{v}_p$ along the surface of
the star $\bm{S}$ and observed that accretion rate varies by different way but on average it does not depend on the
value of the quadrupole component (see Figure 9a). This is an expected result, because the accretion rate should depend
on the viscous properties of the disc but not on magnetic configurations of the  star. All simulations were done at the
same parameters of the disc and fixed value of $\alpha-$viscosity, $\alpha=0.04 $. We should note that initially,
during the first 1-2 rotations accretion rate is very small. This is the time when matter moved towards the disc and
reached the surface of the star through the funnel streams. Later, the structure of the spots establishes and varies
only slightly with time. At present, we are able to simulate accretion to a star with a quadrupole field during $5-50$
rotations. We show the plot in Figure 9 up to $t=5$ because the spots do not change appreciably for longer times.

\begin{figure}
\includegraphics[scale=0.5, angle=0]{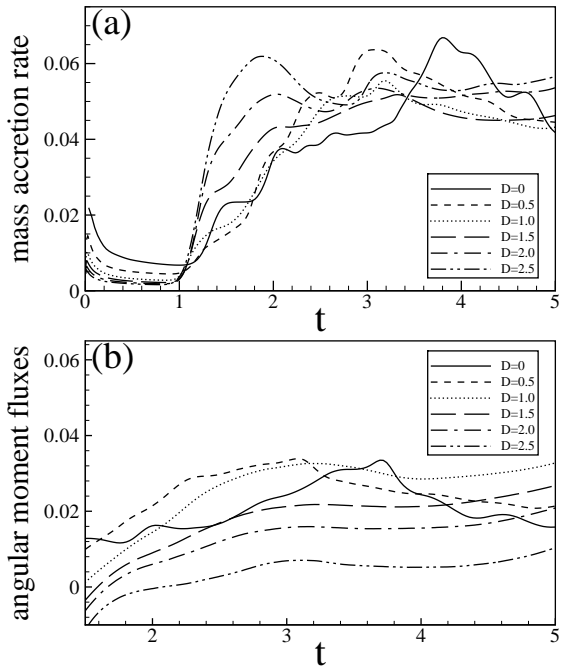}
\caption{\label{fig9} Mass accretion rate $\dot{M}$ (a) and angular momentum fluxes $\dot{L}_f$ (b) for stars with
different strength of quadrupole moment which changes from $D=0$ to $D=2.5$, while $\mu=0.5$, $\Theta=30^\circ$ and
other parameters are fixed.}
\end{figure}

\begin{figure*}
\begin{center}
\includegraphics[angle=0,width=11cm]{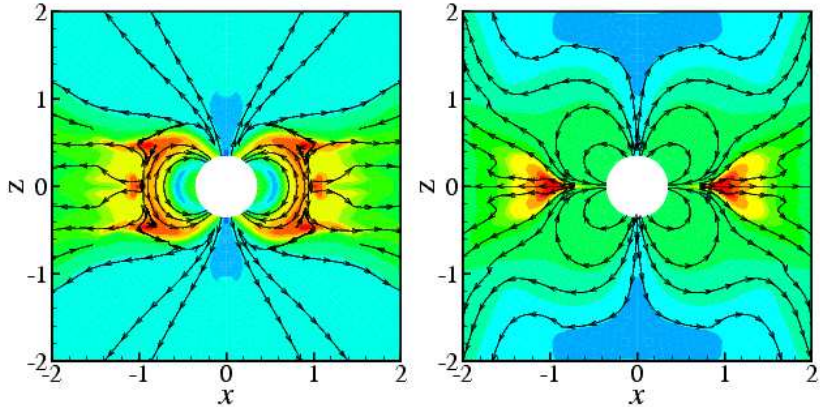}
\caption{\label{fig10} Figure shows streamlines of angular momentum flux carried by the magnetic field lines
$\bm{f_B}=rB_{\phi}\bm{B}_p/4\pi$ in the cases of the pure dipole field  at $\mu=0.5$ (left panel) and pure quadrupole
field $D=0.5$ (right panel) at $\Theta=0$. The background shows the angular velocity  distribution where angular
velocity varies from $\Omega=0.4$ (red) to $\Omega=0.01$ (blue).}
\end{center}
\end{figure*}

We also calculated the spinning torque to the star from the disc-magnetosphere interaction. The main spinning torque
comes from the magnetic field lines connecting a star with the disc and corona $\dot{L}_f=\int d\bm{S}\cdot
rB_{\phi}\bm{B}_p/(4\pi)$ (see also Figures 5 and 6). A star rotates relatively slowly (co-rotation radius is
$r_{cr}=2$) and magnetospheric radius is smaller than co-rotation radius ($r_m\approx 1$) so that the inner regions of
the disc rotates faster than the star and accreting matter spins the star up through the closed field lines connecting
a star with the disc.
%We have a convention that if the torque spins-up the star then it is positive.

There is also a negative torque from the open field lines which start in the polar regions and connect a star with
slowly rotating corona. This torque leads to spinning-down, but this torque is smaller compared to the spin-up torque.
Figure 9b (solid line) shows the torque in the case of the pure dipole star. When we added small quadrupole component,
then torque stayed approximately the same. However, when we added the largest quadrupole component ($D=2$ and 2.5) then
the positive torque decreased significantly. Analysis of angular momentum fluxes in the pure dipole and quadrupole
cases have shown that in the dipole case the main positive torque comes through the inner closed field lines which
connect the disc and the star. Disc rotates faster than the star and angular momentum flows to the star. In the
quadrupole case disc matter flows through the equatorial belt and there is less ``communication" between the star and
the disc. This is why in quadrupole case the positive torque is smaller. This explains the fact that at very large
quadrupole component the spinning torque is smaller: the field structure becomes very similar to a pure quadrupole
topology and the disc accretes through a quadrupole belt.

We investigated the angular momentum flow in greater detail in the cases of pure dipole field ($\mu=0.5$, $D=0$) and
pure quadrupole field ($D=0.5$, $\mu=0$). Figure 10 shows the distribution of the angular velocity (background) and
streamlines of the angular momentum fluxes carried by the magnetic field $\bm{f_B}=rB_ {\phi}\bm{B}_p/4\pi$.  One can
see that in the dipole case, angular momentum flows in from the disk to the star (the star spins-up), and flows out
through the open polar field lines into the corona. In the case of the quadrupole field, angular momentum flows to the
star from the quadrupolar equatorial belt, where the disc matter accretes and is threaded by some closed field lines of
the quadrupole. This is the region where the inward angular momentum flux is the largest. Again, there is angular
momentum outflow from the star through the open polar field lines. In both cases the total angular momentum transferred
to the star is positive and the star spins-up. We should note that the angular momentum carried to the star by matter
is much smaller than that associated with the field, by a factor of 20-100 in the dipole case and 10-20 in the
quadrupole case.

In addition we investigated different properties of the hot spots in cases of aligned ($\Theta=0^\circ$) pure dipole
and pure quadrupole configurations at equal magnetic moments, $\mu=0.5$ and $D=0.5$. In the case of an accreting dipole
the matter of the funnel stream is lifted above the magnetosphere and then is accelerated towards the surface of the
star, so that significant part of gravitational energy is released thus increasing the temperature of the hot spots
(see Romanova et al. 2004). In the case of predominantly quadrupole field matter flows through the equatorial belt
without such dramatic falling to the surface of the star. The simulations show that the velocity of the flow and the
Mach number are about two times smaller in the quadrupole case. The luminosity of hot spots depends on the density and
velocity. The density is similar in both dipole and quadrupole-dominated cases, difference in velocities leads to
conclusion that the hot spots are expected to be $(v_d/v_q)^2$ times dimmer in the quadrupole case. The temperature of
the hot spots is expected to be smaller in the quadrupole case by the factor ($v_d/v_q)^{3/4}$ (about 1.7 times) which
follows from equating of the kinetic energy flux to the black-body radiation flux. In both cases the maximum energy and
highest temperature are in the middle of the spot. Thus, a star with pure dipole case is expected to have hotter and
more luminous spots.

\section{Summary}

We investigated for the first time disc accretion to a star with a dipole plus quadrupole field in full 3D MHD
simulations. The simulations show that for  relatively small misalignment angles $\Theta$, matter flows in a thin wide
sheet to the quadrupole's belt. This flow forms a ring-like hot spot at the surface of the star. The position of the
ring coincides with magnetic equator in the case of the pure quadrupole field, and is displaced into the southern
hemisphere in the case of the dipole plus quadrupole fields. At large $\Theta$, most of matter may accrete to one or
both magnetic poles. In quadrupole case the light curves typically show two peaks per period for the majority of
inclination angles $i$. This might be a sign of the significant quadrupole component. However in the case of pure
dipole field two peaks per period are typical for the case of high inclination of the system (Romanova et al. 2004). If
inclination of the systems is known, then the light curves may help to suggest the possible role of the quadrupole
component.

The simulations show that the torque on the star is larger in the case of the dipole field, where angular momentum is
transferred through the closed field lines connecting the star and the disc. In the quadrupole case the torque is
smaller because matter of the disc accretes directly to the star through the quadrupole belt, that is, between field
lines. There is a positive torque associated with the disc-magnetosphere interaction, however it is much smaller than
in pure dipole case.

In the case of the dominant quadrupole field, the hot spots are expected to be cooler with smaller total energy
released because in the quadrupole belt the funnel flow hits the star with smaller velocities compared to the pure
dipole case. To apply these predictions, one needs to know the accretion rate and the value of the magnetic field on
the surface of the star.

In reality the star's magnetic field may be even more complicated and may include higher order multipoles. In such a
field multiple funnel streams are expected to form (Donati et al. 2006; Jardine et al. 2006). Such multipolar fields
were recently modeled by von Rekowski and Brandenburg (2006) in the axisymmetric time-dependent simulations which
incorporated the dynamo processes of the field formation both in the disc and in the star. In the case of multipolar
fields the light curves are expected to be even more complicated. The light curves obtained in this paper and the light
curves obtained earlier for pure dipole field (Romanova et al. 2004) may help to distinguish the cases where the
magnetic field is highly ordered (dipole or dipole plus quadrupole) from the cases of the multipolar field where the
light curves are expected to be much less ordered. In the future research we plan to investigate the case where the
dipole and quadrupole are misaligned and also the case of accretion to a star with higher order multipolar magnetic
field.

%%%%%%%%%%%%%%%%%%%%%%%%%%%%%%%%%%%%%%%%%%%%%%%%%%%%%%%%%%%%%%%%%

\section*{Acknowledgments} This research was conducted using partly the
resources of the Cornell Theory Center, which receives funding from Cornell University, New York State, federal
agencies, foundations and corporate partners, and partly using the NASA High End Computing Program computing systems,
specifically the Columbia supercomputer. The authors thank A.V. Koldoba and G.V. Ustyugova for the earlier developing
of codes and A. K. Kulkarni for helpful discussions. This work was supported in part by NASA grants NAG 5-13060,
NNG05GL49G and by NSF grants AST-0307817, AST-0507760.

\section*{Appendix}

We perform full 3D MHD numerical simulations. Equations are written in a reference frame rotating with the star, with
the $z-$axis aligned with the star's rotation axis,
\begin{equation}
\frac{\partial\rho}{\partial t}+\nabla\cdot{(\rho\bm{v})}=0 \ \ ,
\end{equation}
\begin{equation}
\frac{\partial(\rho\bm{v})}{\partial t}+\nabla\cdot\bm{T}=\rho\bm{g}+
2\rho\bm{v}\times\bm{\Omega}-\rho\bm{\Omega}\times(\bm{\Omega}\times\bm{R}) \ \ ,
\end{equation}
\begin{equation}
\frac{\partial(\rho S)}{\partial t}+\nabla\cdot(\rho S\bm{v})=0 \ \ ,
\end{equation}
\begin{equation}
\frac{\partial\bm{B}}{\partial t}-\nabla\times(\bm{v}\times\bm{B})=0 \ \ ,
\end{equation}
%\[\frac{\partial\rho}{\partial t}+\nabla\cdot{(\rho\bm{v})}=0,\]
%\[\frac{\partial(\rho\bm{v})}{\partial t}+\nabla\cdot\bm{T}=\rho\bm{g}+
%2\rho\bm{v}\times\bm{\Omega}-\rho\bm{\Omega}\times(\bm{\Omega}\times\bm{R}),\]
%\[\frac{\partial(\rho S)}{\partial t}+\nabla\cdot(\rho S\bm{v})=0,\]
%\[\frac{\partial\bm{B}}{\partial t}-\nabla\times(\bm{v}\times\bm{B})=0,\]
where $\bm{v}$ and $\bm{B}$ are vectors of velocity and magnetic field in the three-dimensional space, $\rho $ is
density, $\bm{\Omega}$ is the angular velocity of rotation of the star, $S$ is the entropy per gram, $\bm{g}$ is the
gravitational acceleration, $\bm{T}$ is the stress tensor. Using the reference frame rotating with the star we
eliminate from equation large terms corresponding to the magnetic forces induced by rotation of the strong field
associated with connected with rotation

The problem of accretion to a star with a dipole and quadrupole magnetic fields is complicated because of high
gradients of the magnetic field, $B_d\sim 1/R^3$ for the dipole and $B_q\sim 1/R^4$ for the quadrupole components. To
avoid this difficulty and other difficulties connected with strong magnetic field near the star, the magnetic field was
decomposed into the ``main" component, which is fixed and consists of the dipole and quadrupole parts:
$\bm{B}_0=\bm{B}_d+\bm{B}_q$, and variable component $\bm{B_1}$ which is induced by currents in the simulation region
and is calculated in equations. This splitting helped to decrease the  magnetic force (Tanaka 1994; Powell et al.
1999).

A special ``cubed" sphere grid was developed for solution of such a problem which has advantages of the spherical and
Cartesian coordinate systems (Koldoba, et al. 2002). A grid consists of a sequence of co-centric spheres (cubes
inflated to the sphere), which are placed at different radii $R$ from the center of coordinates (where a star is
located). At each inflated cubed sphere there are six sectors corresponding to the sides of the cube, and curvilinear
Cartesian coordinated are introduced in each of six sides with $N^2$ cells. The concentric ``spheres" are
inhomogeneously distributed such that the grid is very fine closer to the star and it is much coarser far from the
star.  Each grid cell has approximately the same sides in all three directions and this corresponds to refine grid near
the star. Thus, the whole simulation region consists of six blocks with $N_R\times N^2$ cells. In the current
simulations we took the grid $75\times31^2$ in each of six blocks. Other grids were investigated for comparison. The
coarser grids give satisfactory results in the case of the pure dipole field, however in the case of the quadrupole
field, the code requires finer grid, because of the higher magnetic field gradients. The 3D MHD code was parallelized,
so that the jobs can be run on $N=6\times n$ processors. At $N>60$ the communication time between $R-$regions becomes
significant, but in the future in the case of finer grid, we can use even larger number of processors.

To solve the above partial differential equations, we used a high order Godunov-type numerical code earlier developed
in our group (Koldoba et al. 2002). The numerical scheme is similar to one described by Powell et al. (1999).  The
Godunov-type schemes have been recently well described in the literature (see, e.g. book by Toro 1999).  In addition we
incorporated viscosity to the numerical scheme to control the rate of matter flow to the star. The main goal of
viscosity is to bring matter from the far regions of the disc to the star, so that we used a simplified $\alpha-$
prescription with $\alpha=0.04$ in all simulation runs with smaller/larger values for testing.

\end{document}